# Future Illiteracies—Architectural Epistemology and Artificial Intelligence

Mustapha El Moussaoui

Faculty of Design and Art, Free University of Bolzano, 39100 Bolzano, Italy; mustapha.elmoussaoui@unibz.it

**Abstract**

In the age of artificial intelligence (AI), architectural practice faces a paradox of immense potential and creeping standardization. As humans are increasingly relying on AI-generated outputs, architecture risks becoming a spectacle of repetition—a shuffling of data that neither truly innovates nor progresses vertically in creative depth. This paper explores the critical role of data in AI systems, scrutinizing the training datasets that form the basis of AI's generative capabilities and the implications for architectural practice. We argue that when architects approach AI passively, without actively engaging their own creative and critical faculties, they risk becoming passive users locked in an endless loop of horizontal expansion without meaningful vertical growth. By examining the epistemology of architecture in the AI age, this paper calls for a paradigm where AI serves as a tool for vertical and horizontal growth, contingent on human creativity and agency. Only by mastering this dynamic relationship can architects avoid the trap of passive, standardized design and unlock the true potential of AI.

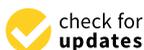

## 1. Introduction

The emergence of artificial intelligence (AI) has brought about a profound transformation across numerous disciplines, reshaping established norms, practices, and epistemologies. In architecture, the allure of AI-driven design lies in its ability to generate vast outputs, simulate complex processes, and produce visually striking forms with unprecedented speed and precision. Yet, beneath this surface of spectacle and technological marvel, critical questions emerge: What happens when architects become passive consumers of AI-driven outputs? How does reliance on machine-generated designs reshape architectural knowledge, creativity, and practice? And what is lost when machines, rather than human minds, become the primary agents of knowledge creation and reproduction?

This paper argues that AI, while a powerful and potentially transformative tool, risks reducing architecture to a state of standardized repetition when used uncritically. AI systems, driven by training data and algorithms, inherently reflect the biases, values, and limitations embedded within their datasets. Left unchecked, they produce architectural outcomes that mirror and recombine existing patterns, reinforcing rather than challenging dominant norms. The result is a horizontal expansion of knowledge—a broad proliferation of design possibilities characterized by superficial variation. In contrast, vertical growth is conceptualized as profound and transformative development, driven by reflective practice and contextual insight.

Byung-Chul Han's analysis of the digital condition reveals how algorithmic systems erase depth from thought by collapsing the subject into a series of measurable performances.

In such a flattened world, epistemology degenerates into logistics; optimized, accelerated, and stripped of interiority [1]. Architectural creativity becomes a byproduct of machinic sequencing, rather than an emergent act of human interpretation. Additionally, Kate Crawford and Trevor Paglen, in their examination of the machinic unconscious, expose how layers of training data and algorithmic filtration obscure the logic of image generation behind a veil of operational abstraction [2]. Architectural tools powered by these models do not simply "generate images"; they encode histories of exclusion, aesthetic preference, and normalized bias. To deploy such systems without epistemological reflexivity is to inherit the blindness of the dataset. This terrain is further examined by Bratton and Parisi, arguing that AI constitutes a form of alien reasoning that outpaces human cognition not only in capacity but in form [3,4]. The stack of planetary computation reconfigures the way knowledge is synthesized, demanding that human actors re-situate themselves within a vastly expanded epistemic ecology. For architecture, this means reconceiving design not as a linear act of authorship but as a strategic intervention within layered, contingent, and computationally mediated systems.

At the heart of this challenge lies the question of human agency. Passive reliance on AI-driven tools transforms architects into operators of algorithmic engines rather than authors of meaning. This leads to a flattening of architectural knowledge into endless permutations. Without active epistemological intervention—vertical layering—knowledge remains shallow and context-free. To navigate this tension, architects must reclaim their roles as active creators and critical thinkers. It is through the deliberate interruption of machinic flow, the reassertion of interpretive agency, and the critical examination of data provenance that architectural epistemology can be revived rather than replaced.

Underpinning this article is a recognized knowledge gap: While technical debates about AI tools are rich, the epistemological implications of passivity versus critical engagement remain under-theorized in architecture. To address this, a qualitative methodology comprising literature artifact and content analysis, investigating student-generated and AI-assisted design outcomes to reveal how epistemic depth is enacted—or omitted. This paper then explores conditions for epistemological layering—from data, creativity, and critique—and argues for a paradigm of integrated practice where AI's capacity for breadth complements human reflexivity and contextual wisdom.

## 2. Perspectives on Architectural Knowledge

Architectural knowledge has long been a multifaceted construct, shaped by both theoretical and practical dimensions. Vitruvius, in his De architectura, emphasized the triad of firmitas, utilitas, and venustas (firmness, utility, and beauty) as foundational pillars of architectural practice [5]. This conceptual framework underscored the necessity of harmonizing structural integrity with functional and aesthetic considerations, a vision that has remained a touchstone in the epistemological evolution of the discipline.

The Enlightenment era brought with it a wave of rationalism, wherein architecture was redefined as a system of knowledge bound by principles of reason, symmetry, and order [6]. Through this lens, architecture was not merely a craft but a mode of intellectual inquiry, integrating philosophy, mathematics, and the arts. This tradition of synthesis established a basis for understanding architectural knowledge as contingent upon both cognitive and sensory faculties.

In more recent times, architectural theory has grappled with the implications of modernism and postmodernism, wherein epistemological frameworks have been subject to intense critique and reformulation. Scholars like Tafuri have argued for a critical historiography that resists reductionist readings, emphasizing the socio-political entanglements within which architectural knowledge is produced [7]. Such a perspective illustrates that architectural epistemologies are not static; they are dynamic and evolve in response to broader cultural, technological, and philosophical currents.

Recent scholarship has deepened this analysis, particularly in the context of artificial intelligence. Some studies highlighted that the rise of AI demands a reevaluation of tacit and explicit knowledge [8,9]. While explicit knowledge—codified principles and formal techniques—can be efficiently processed by AI systems, tacit knowledge remains embedded in human intuition, experiential judgment, and embodied cultural awareness. This duality is critical because human contextual understanding is essential for interpreting AI's outputs meaningfully, especially when navigating local or site-specific design challenges.

Recent neuroscientific evidence reinforces the epistemological concerns raised by the overreliance on AI in knowledge production. A comprehensive study assessed the neural and behavioral consequences of using large language models (LLMs) like ChatGPT 3.5 in essay writing tasks. Utilizing EEG, NLP, and human–AI evaluations, the authors demonstrated that participants who consistently used an LLM showed significantly lower cognitive engagement, weaker brain connectivity, and reduced alpha and beta activity—key indicators of focus and memory processing—compared to those relying on their own faculties or using a search engine (Figure 1) [10].

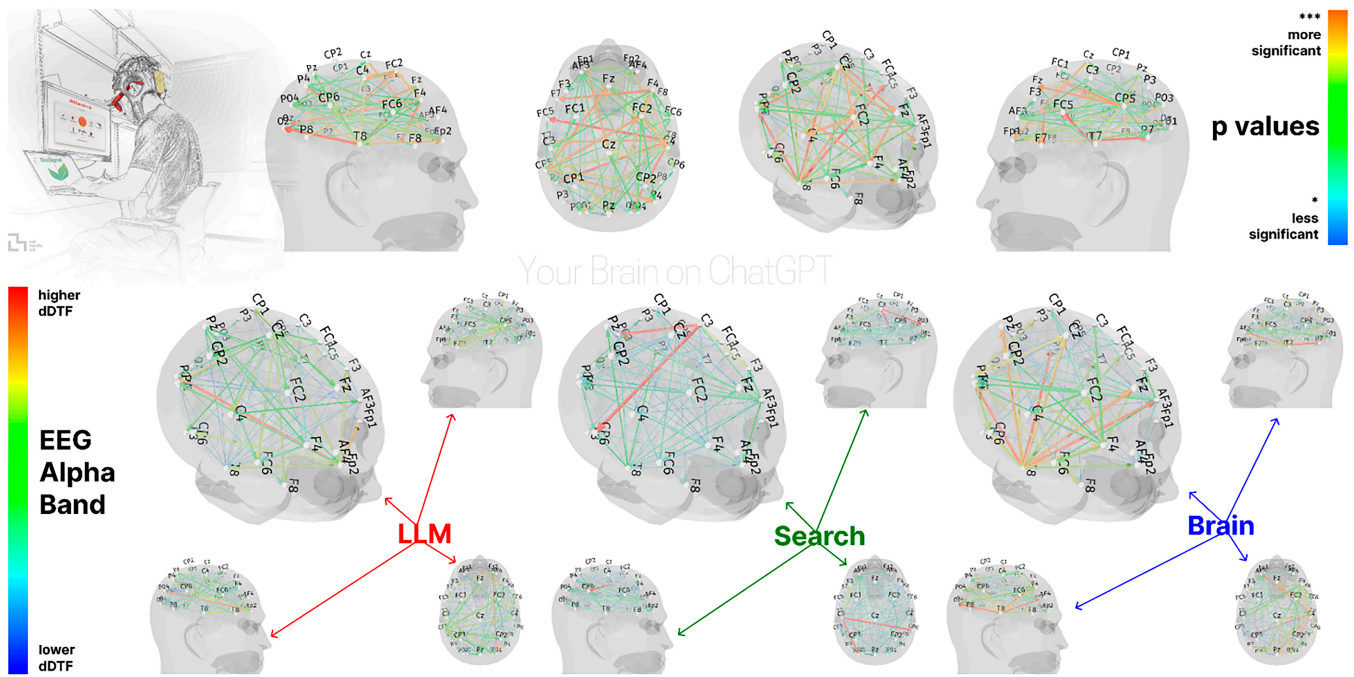

**Figure 1.** The dynamic Direct Transfer Function (dDTF) EEG analysis of Alpha Band for groups: Search Engine, Brain-only, LLM, with *p*-values to demonstrate significance from moderately significant (*) to highly significant (***) [10].

With that being said, likewise, architectural epistemology faces new challenges and opportunities with the advent of AI. AI technologies, particularly generative AI systems capable of autonomously proposing architectural forms, introduce profound epistemological shifts. Central to understanding this transformation is the distinction between explicit and tacit knowledge in architecture. Explicit knowledge involves codifiable rules such as geometric principles and structural logic, easily translated into data for AI systems. It is the kind of knowledge that can be captured, formalized, and input into computational models. Conversely, tacit knowledge refers to the non-codified, intuitive, and experiential knowledge architects develop through practice. It is embedded in bodily perception, sensory engagement, historical memory, and cultural literacy that comes from immersion in place and practice. This form of knowledge cannot be easily verbalized or digitized and is often exercised in moments of aesthetic judgment, spatial interpretation, or creative divergence. Thus, while AI may simulate outputs based on prior data, it cannot truly "know" in the sense that human architects do—particularly in conditions of novelty, ambiguity, or ethical tension. Consequently, human oversight and creative intuition are essential to effectively leverage AI in architecture.

Moreover, contemporary theoretical contributions frame AI as a co-creative agent, emphasizing the potential for human–machine collaboration to expand the architectural design space. While AI enhances designers' capacity for rapid iteration and novel exploration, scholars mentioned are advocating instead for hybrid intelligence models that balance machine-driven suggestions with human criticality and originality [11,12].

The cognitive process underpinning architectural creativity can be visualized as a problem-solution dynamic involving continuous iteration between conceptualization and representation (Figure 2). This framework begins not with form but with intellectual engagement—with the ethical, cultural, historical, and philosophical dimensions of the problem. Through critical thinking and imagination, the architect draws upon tacit knowledge to shape the conceptual phase. Although depicted simplistically, this process is nonlinear and complex, involving simultaneous engagement with philosophical, cultural, historical, and technical knowledge, culminating in a creative architectural solution. Once this mental conceptualization occurs, it is externalized through representational techniques such as descriptive geometry, CAD software, and visualization tools.

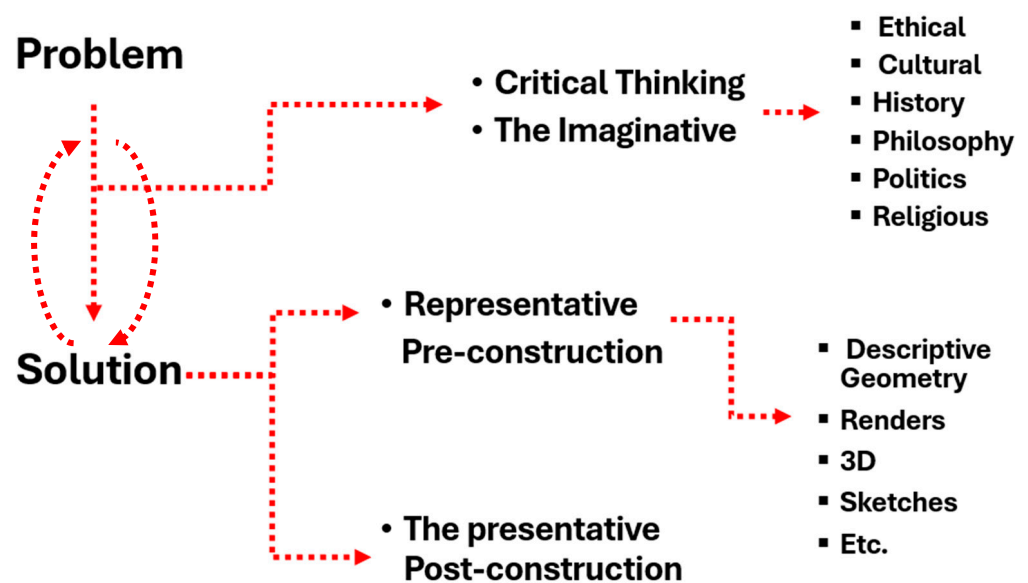

**Figure 2.** Conceptual Problem-Solution Pathway: Visualizing the layering of tacit and explicit knowledge. Source: Author.

Figure 2 underscores the critical role of tacit knowledge in linking the problem to the solution. The upper pathway highlights the intangible yet foundational dimension of creativity—rooted in cultural and philosophical reflection—while the lower path reflects the externalizations AI excels at. This bifurcation reinforces that true architectural knowledge emerges from the dynamic interplay of these levels.

Currently, AI excels significantly in representational tasks, outperforming human capabilities in producing rapid, detailed, and visually compelling outputs. However, in the critical phase of conceptualization and creative ideation, AI remains limited due to its inherent reliance on previously analyzed data. It lacks the intuitive capacity and contextual understanding necessary to independently generate genuinely innovative design solutions, as creative ideation involves navigating complex and multifaceted variables beyond mere data analysis [13].

Recent literature suggests that AI's impact on architectural education is profound but double-edged [12,14–16]. On the one hand, AI offers new pathways to formalize and test design principles, potentially overcoming the subjectivity that has traditionally dominated architectural discourse. On the other hand, scholars warn that AI's outputs are shaped by the biases embedded in its data and algorithms, making it essential for architects to remain vigilant in curating and interpreting these outputs. The future of architectural knowledge lies in integrating empirical, data-grounded insights with the qualitative, humanistic knowledge that defines the field. This balanced integration will ensure that architecture continues to evolve as both an evidence-based and culturally grounded discipline.

## 3. The Role of Data in AI Systems

At the core of every artificial intelligence (AI) system lies data—a raw, malleable substance from which all machine intelligence is derived. The process of data insertion, commonly referred to as training, involves feeding vast quantities of data into algorithms to enable them to identify patterns, recognize trends, and generate outputs that mimic creative or analytical processes. The nature and quality of this data are paramount; they determine not only what AI can do but also what it cannot [17]. In architectural applications, this dependency manifests in AI's propensity to emulate existing design paradigms, rooted in datasets of past works. Thus, AI's creative potential is constrained by the scope of the data on which it is trained.

The selection of data is never neutral. Training datasets reflect specific aesthetic, cultural, and historical biases that shape AI's outputs [18]. For example, if a dataset predominantly contains works from a specific architectural movement, the AI model may disproportionately favor its stylistic traits, potentially marginalizing other architectural vocabularies. This process, referred to as "data inheritance," describes how an AI system's outputs are deeply shaped by the epistemological traces embedded in the datasets it consumes, meaning that the AI inherits not just raw information but also the underlying worldviews, stylistic patterns, and cultural assumptions of the data's origins. Clarifying this term is essential because it highlights the AI's dependence on prior human choices, framing it as an epistemological relay rather than a neutral technical process.

As a result, the architectural outputs of AI models often reflect existing hegemonies and fail to challenge or innovate beyond them. This risk is compounded by the opacity of many AI systems, wherein the inner workings of the model—what it learns and why—remain hidden [19]. For architects, this raises critical questions: Who selects the data? What criteria are used to include or exclude particular works? And how does this process shape the built environment of the future?

Training data for AI systems are often curated, cleaned, and annotated to ensure that models learn effectively. This data-driven process transforms raw inputs into structured knowledge that algorithms can process, classify, and reproduce. In architectural practice, these datasets may include images, schematics, structural calculations, and textual descriptions, which collectively provide a broad representation of the discipline. However, the process is far from infallible. The potential for AI to homogenize architectural practice poses a significant epistemological risk. By relying on training datasets derived from existing works, AI systems often reinforce prevailing architectural norms and biases. This phenomenon, described by Bourdieu as the "habitus" of reproduction, can lead to a cycle of standardization wherein novel architectural expressions are subsumed by pre-existing patterns [20].

As AI systems generate designs based on the statistical aggregation of past works, they risk privileging repetition over innovation—a reality that challenges architecture's creative ethos. For example, diffusion models like those utilized in ComfyUI represent a significant advancement in how data are integrated and synthesized to generate outputs. Diffusion models operate by progressively refining noise into coherent forms, guided by a learned probability distribution derived from training data [21]. Unlike traditional generative models, diffusion frameworks iterate over a series of steps, using a reverse diffusion process to reconstruct architectural imagery or design elements from latent spaces. This mechanism relies heavily on the quality and diversity of the datasets provided during training, as the model learns to "denoise" by identifying recurring patterns and correlations.

The types of data included in such systems are vast and varied, often spanning images, 3D models, construction drawings, material properties, and textual annotations. For example, in ComfyUI, users can customize their workflows by introducing pre-curated datasets or augmenting existing ones with personalized content.

For instance, in ComfyUI, one can utilize pretrained models readily available on various online platforms. As illustrated in Figure 3, the model showcased—an interior design model trained by Sa_May—has been exclusively trained on the aesthetic category of Bionic Futuristic Interior Design. Consequently, regardless of the input sketch or textual prompt, the generated outcome will always be constrained by the visual logic embedded within this pretrained dataset. While such tools are technically impressive, they risk producing designs locked within narrow stylistic confines, amplifying the standardization effect already discussed. Without critical human intervention, the use of pretrained models like those in ComfyUI reinforces the bias toward familiar patterns rather than fostering genuine innovation, embedding the same aesthetic tropes deeper into the architectural landscape.

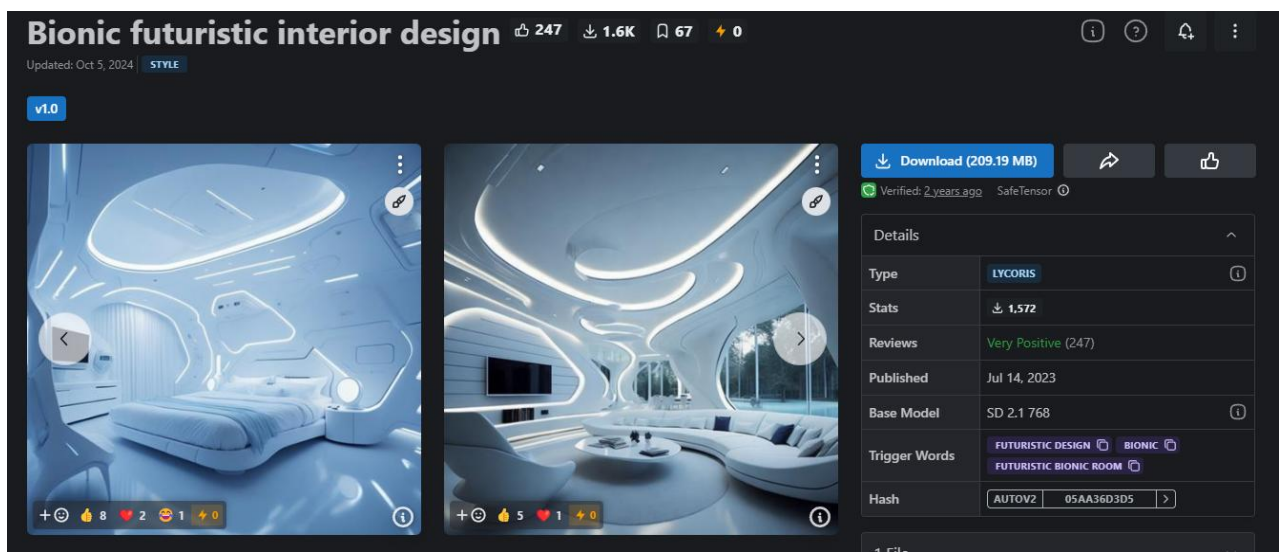

**Figure 3.** Pretrained model developed by Sa_May on Bionic Futuristic Interior Design.

## 4. Biases and Limitations in AI Training

AI systems are inherently biased, not by virtue of their programming but by the data from which they learn [18]. This bias is particularly pronounced in architecture, a field steeped in cultural symbolism and historical context. As we have seen earlier, when datasets are skewed toward modern-style architectural traditions, for instance, AI models may privilege gothic forms and styles, relegating other style expressions to the periphery. This could be due to the trainset of the module used for the ComfyUI, for example.

Moreover, AI systems trained on static datasets struggle to adapt to changing sociocultural contexts [22], as architecture, by its nature, evolves with the needs and aspirations of society. If AI models are trained on previous data, they risk producing designs that fail to respond to contemporary challenges. This temporal rigidity highlights the limitations of data-driven design and underscores the need for dynamic and continually updated datasets that reflect architecture's fluid nature.

Due to the immense influence that data exert over AI systems, architects and AI practitioners bear a profound ethical responsibility. The curation of training data is not a neutral act; it shapes AI's understanding of what architecture "is" and what it "should" be. This power demands critical reflection and accountability. Without proper oversight, AI can become a tool for perpetuating existing inequalities and biases, undermining architecture's potential as a vehicle for societal transformation. Ethical data curation involves actively seeking diverse and inclusive datasets, challenging dominant narratives, and foregrounding historically underrepresented perspectives [23].

## 5. The Cycle of Machine-Driven Consumption and Reproduction

The processes of reading, writing, and reproducing knowledge have become deeply entwined with computational mechanisms. The implications of this transformation extend beyond mere automation; they signal a paradigm shift in which machines not only generate outputs but also consume and reproduce them. In architecture, this phenomenon is exemplified by the proliferation of generative adversarial networks (GANs) and diffusion models, which produce highly detailed, often spectacular images that serve as the basis for further AI-generated iterations [24].

This cycle of consumption and reproduction results in what Baudrillard termed the "hyperreal," a condition in which simulations or representations detach from their original context and reality, becoming self-sustaining entities that replace authentic experiences and truths with convincing, yet ultimately superficial, images [25]. Translated into architectural epistemology, the hyperreal manifests when AI-generated forms no longer refer meaningfully to original architectural ideas, contexts, or cultural narratives but instead continuously reference and replicate earlier simulations, increasingly detached from reality.

The capacity of AI systems to "hallucinate" images—generating forms based on probabilistic mappings of data—amplifies the challenge of horizontal knowledge reproduction [26]. This process, while capable of producing visually striking outputs, remains fundamentally rooted in the recombination of pre-existing data. The generative process is constrained by the limitations and biases of its training datasets, leading to outputs that reflect, amplify, and occasionally distort the norms encoded within the data [27].

Architecture, as a discipline, is particularly vulnerable to this phenomenon. When AI-generated designs are consumed and reproduced without critical intervention, they perpetuate a cycle of aesthetic and conceptual repetition, especially when these tools are used by non-trained architects/designers (even non-mature architects/designers), which flattens the output into a spectacle rather than being capable of solving a deeper fundamental issue. This leads to a flattening of architectural knowledge, wherein the emphasis shifts from innovation to the endless reproduction of AI-driven spectacles. As Han observes, such a state of passive consumption risks reducing creative engagement to the mere act of viewing, stripping away the critical and transformative potential of architectural practice [28].

AI outputs, when left unchecked, propagate a form of horizontal knowledge growth. This growth is characterized by an expansive breadth of possibilities derived from existing data without a corresponding depth of critical or creative engagement (Figure 4). As AI recombines and reinterprets architectural forms, it can spread knowledge laterally, exploring permutations within established parameters. However, this horizontal expansion, while vast, lacks the transformative power of vertical growth, the deepening of knowledge through critical, contextual, and creative input. Importantly, "epistemological layering" differs from conventional design iteration by intentionally integrating multiple domains of knowledge (historical, ethical, cultural, philosophical) into each step, rather than merely refining form or function across successive versions. While iteration focuses on incremental adjustments, epistemological layering emphasizes cross-disciplinary depth and conceptual integration.

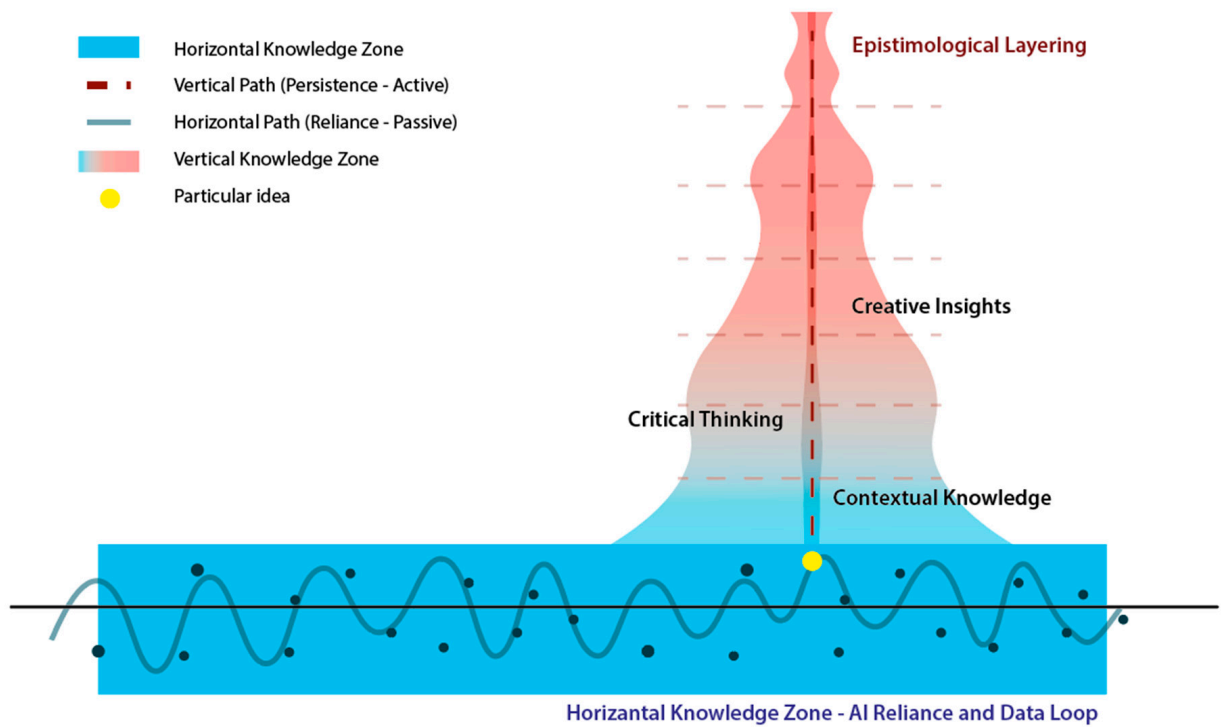

**Figure 4.** Epistemological layering–Horizontal vs. Vertical Knowledge Source: Author (2025).

The concept of horizontal growth and the AI connecting possibilities, initially, could allow us to imagine it like "rhizome," a structure characterized by decentralized, lateral connections without clear hierarchical roots [29]. However, while Deleuze and Guattari's rhizome symbolizes positive multiplicity, richness, and productive interconnectedness, AI-driven horizontal growth in architecture does not necessarily embody these qualities. Instead, it often leads to superficial, repetitive iterations devoid of critical reflection, innovation, or deeper meaning.

In contrast, vertical epistemological growth necessitates the active and critical intervention of architects. This active stance involves consciously breaking free from data-driven cycles through reflective and creative processes, a practice referred to here as "epistemological layering." Epistemological layering is the deliberate process through which architects integrate multiple levels of knowledge, historical, cultural, contextual, technological, philosophical, etc., to produce innovative solutions. Rather than relying solely on AI-generated outputs, architects who engage in epistemological layering critically interrogate, transform, and transcend these outputs, deepening architectural understanding through each successive iteration. This conceptual framework is visualized in Figure 4, which presents the dynamic oscillation between horizontal and vertical knowledge expansion as a fluid epistemological continuum. To further situate this model within a dialectical matrix, a complementary quadrant diagram (Figure 5) is introduced. The quadrant diagram categorizes the epistemic conditions of architectural design across varying levels of agency and creativity. While Figure 4 emphasizes the temporality and internal rhythm of knowledge deepening, Figure 5 enables critical positioning and argumentative clarity by mapping the epistemic consequences of AI–human interaction across four quadrants. The dual inclusion of these diagrams reflects the necessity of both ontogenetic flow and topological distinction in the analysis of AI's impact on architectural knowledge.

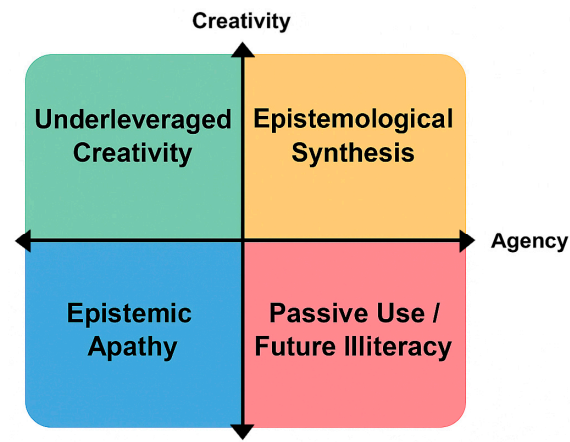

**Figure 5.** Epistemological Conditions of AI-Driven Design: A quadrant model mapping the intersection of creativity and agency. The horizontal axis represents the degree of agency (from low to high), while the vertical axis represents creativity (from low to high). Source: Author (2025).

For instance, an architect can critically engage with AI-generated designs by rigorously assessing their contextual appropriateness, historical significance, and social implications, thereby enriching them with layers of critical insight. Alternatively, architects can integrate extensive site-specific data, including location, orientation, regulations, and zoning laws, into the AI system, allowing it to analyze this information at super speed and generate a wide range of design possibilities. Through such an approach, what initially begins as a superficial AI output is refined into a deeply contextualized and meaningful architectural solution, thereby achieving genuine vertical epistemological growth.

To integrate AI-driven computation with human creativity, utilizing epistemological layering while leveraging the vast expanse of knowledge available in the horizontal knowledge zone, my teaching approach in AI and Architectural Design emphasizes critical engagement with AI tools while preserving creative autonomy (Figure 6).

Students are guided through a structured yet open-ended iterative design process—beginning with hand-drawn sketches, progressing through analytical and conceptual development, and culminating in refined outputs generated through AI-driven tools. This dialogue between the analog and the digital encourages students to understand AI not as a substitute for creativity, but as a catalyst and amplifier of their design intentions.

The course, titled AI and Architecture: From Theory to Practice, is offered at the Free University of Bozen-Bolzano within the Studium Generale framework (10 students). As an elective open to students across all academic levels—from undergraduate to doctoral—it fosters interdisciplinary exchange and diverse engagement with AI in design. A condensed version of this course was also delivered as a 10-h intensive workshop at Brno University of Technology, targeting primarily Master's students and final-year Bachelor's students in architecture (25 students).

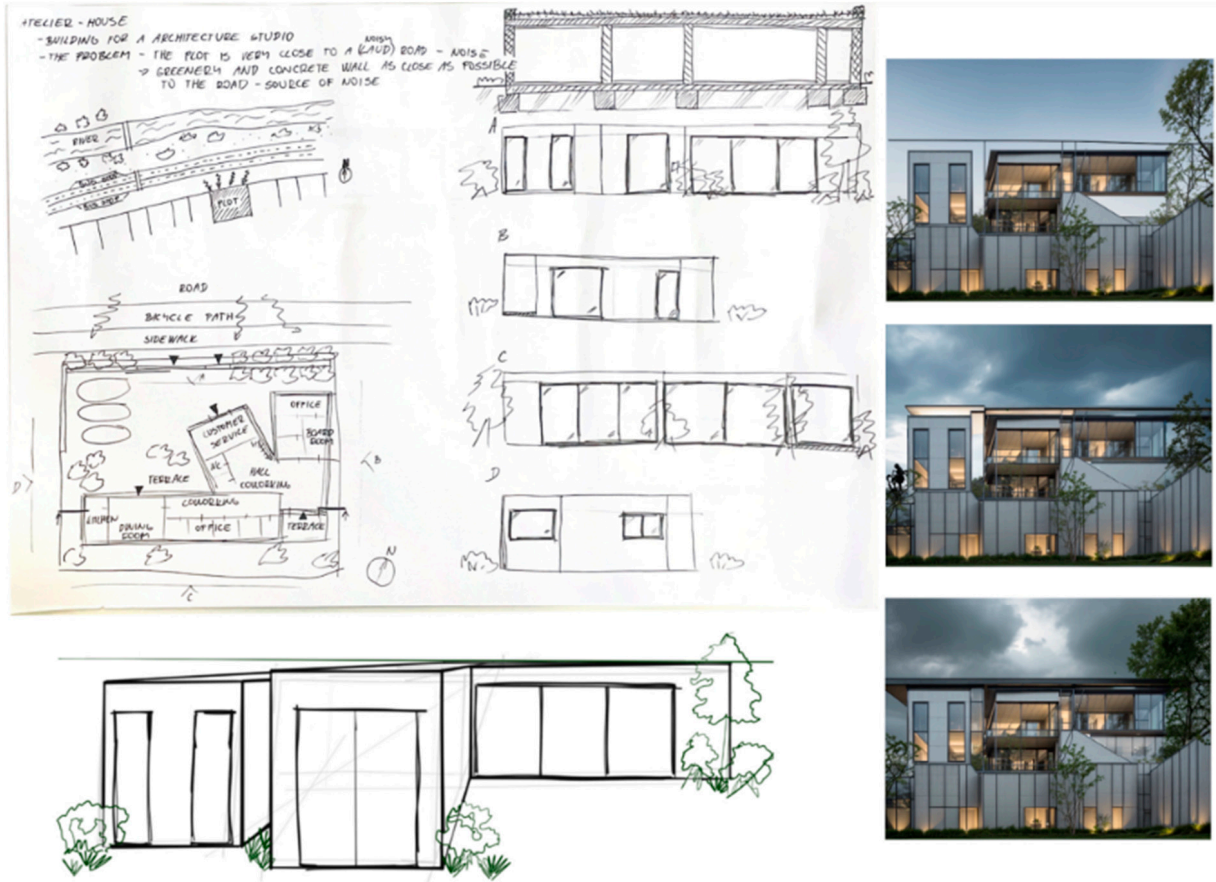

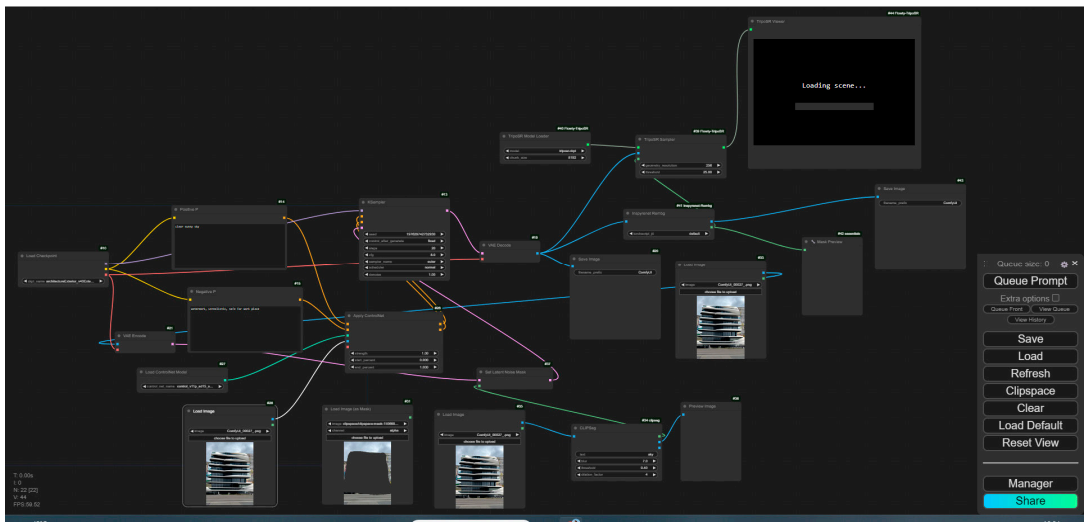

**Figure 6.** Students at Brno University using ComfyUI AI tool to control the machine, part of the Authors AI and Architecture course. Authors: Iveta Rohanova & Gajdova Gabriela.

Throughout the course, students are invited to select a site of personal or architectural relevance and to define a design problem. They begin by formulating initial conceptual approaches through sketching and diagramming. Subsequently, they are introduced to the ComfyUI platform, where they explore visual generation via a nodal interface. This process involves creating iterative representational sequences by linking functional nodes, adjusting prompts, and introducing the initial sketches to the latent space through Controlnet. As their AI-generated representations evolve, students are encouraged to return to and revise their original sketches and conceptual frameworks, establishing a recursive loop between intention and output, abstraction and visualization. This approach allows for an active, critical stance, encouraging students to apply their intellectual capacities and personal perspectives to architectural challenges while allowing AI to handle repetitive or computationally intensive tasks. AI thus functions as a highly capable assistant—an "assistant on steroids"—enabling architects to reallocate time toward deeper intellectual pursuits, conceptual refinement, and strategic design thinking. In the course, students successfully transformed a hand-drawn sketch into a 3D obj model, ready for 3D printing, in under 18 teaching hours using an open-source tool. This tool facilitates rapid prototyping but also allows students to modify, program, iterate, and train their own AI models, reinforcing the idea that AI should be an adaptable, customizable extension of the designer's creative vision.

## 6. Conclusions—The Co-Evolution of Human Creativity and AI

As artificial intelligence (AI) becomes increasingly central to architectural practice, a crucial epistemological challenge emerges regarding the interplay between machine-driven processes and human creativity. The architectural discipline finds itself at a pivotal juncture, navigating between passive acceptance of AI-generated outcomes and active, critical engagement that preserves the profession's intellectual rigor and creative depth. This challenge requires architects to consider carefully the conditions under which human creativity and machine intelligence might co-evolve beneficially, transcending mere automation to achieve genuinely innovative and transformative design practices.

The notion of a symbiotic or co-evolutionary relationship between human creativity and AI is founded upon the recognition that each holds unique epistemological strengths and limitations [30]. AI excels at processing vast quantities of explicit, structured data, rapidly generating a wide array of permutations and options with remarkable speed and precision. Yet, despite this impressive computational capability, AI remains fundamentally constrained in its capacity for genuine creative intuition, contextual sensitivity, ethical reflection, and tacit knowledge—all qualities inherently human. Human creativity, in contrast, draws upon tacit knowledge rooted in lived experience, sensory perception, intuition, and deep cultural understanding—qualities that remain elusive to machine intelligence [31]. Thus, the creative capacity of humans remains indispensable, particularly when addressing complex, contextually nuanced architectural challenges.

For architects, the essential task is not simply to use AI as a tool but to actively shape and interrogate its outputs. Through such active engagement, AI becomes not just a generative mechanism but an epistemological partner, extending human reach without replacing human judgment. This active interplay enriches architectural knowledge both horizontally, by exploring diverse aesthetic possibilities, and vertically, by delving into deeper layers of critical, cultural, and philosophical insight.

Importantly, "epistemological integration" differs from the earlier models of mere iteration or layering. While epistemological layering refers to enriching a design with added historical, cultural, or contextual dimensions, epistemological integration denotes a full fusion of human and machine—an intertwining where critical human agency and machine computation are mutually shaping, forming a co-evolutionary intelligence greater than either alone. It is not about adding more layers; it is about restructuring the design process itself so that human reflective capacities and AI-driven permutations function in constant dialectical tension.

This perspective aligns with Martin Heidegger's view on technology, particularly his notion of Gestell, or enframing. Heidegger warned that when technology is adopted uncritically, it reduces humans to mere operatives of a system, stripping away reflection and agency. In architecture, passive reliance on AI similarly risks diminishing the architect's role, making them custodians of algorithmic outputs rather than authors of innovation. Heidegger argued that true mastery over technology requires not rejection, but active, critical awareness—a stance in which human creators consciously shape, direct, and challenge technological processes [32].

Practically, achieving epistemological integration requires rethinking both architectural education and practice. Architects must be trained not only to operate AI systems but to question how these systems shape knowledge, aesthetics, and social values. Curricula must embed critical, ethical, and philosophical inquiry alongside technical training, ensuring that human intentionality remains at the center. In professional practice, architects must construct workflows that balance AI's rapid generative power with the slower, more deliberative processes of human judgment, intuition, and contextual understanding.

A concrete example of epistemological integration can be seen when an architect critically evaluates an AI-generated design proposal. Rather than accepting the AI output as final, the architect interrogates its social appropriateness, cultural relevance, and spatial sensitivity, then reshapes it using insights drawn from experience, intuition, and contextual expertise. The resulting design becomes a hybrid outcome—not purely human nor purely machine, but a synthesis born from an active and ongoing negotiation between both.

This transition of approach towards architecture brings us directly to the concept of the "future illiterate," a term that captures the stark consequences of failing to integrate critically with technological systems. Thus, passive submission to AI's overwhelming outputs, combined with uncritical acceptance of its designs, leads to a condition where the architect ceases to think beyond the preconfigured parameters of the system. Trapped within a self-referential circuit of machine-generated knowledge, the future illiterate no longer exercises discernment, no longer innovates, but merely consumes and perpetuates what has already been inscribed. This closing warning is not simply rhetorical; it is a call to reassert the human force of will, to resist passivity, to maintain sovereign agency against the inertia of automation, and to ensure that the creative spirit of architecture endures within the AI age.

In conclusion, this paper has argued that architectural practice stands at a decisive epistemological crossroads. To avoid the flattening effects of passive AI consumption, architects must embrace an active, critically integrated stance, leveraging the strengths of both human creativity and machine computation. By doing so, they can cultivate a future in which AI augments rather than diminishes architectural knowledge, producing designs that are not only technically sophisticated but also culturally meaningful, contextually sensitive, and profoundly human. Future research should explore how educational curricula, design workflows, and professional ethics can be reshaped to support this integrated paradigm, ensuring that architects remain not passive consumers, but active shapers of the future built environment.

**Funding:** Research Funded by the European Social Fund Plus Project code ESF2_f3_0006 CUP: B56F24000090001. Research projects on strategic topics for the South Tyrolean economy carried out by researchers on fixedterm contracts. The work was also supported by the Open Access Publishing Fund of the Free University of Bozen-Bolzano (which covered the open access fee).

**Institutional Review Board Statement:** Ethical review and approval were waived for this study due to the fact that the nature of the study did not involve sensitive personal data or interventions. What is shown is a screenshot of a program (ComfyUI) while they were through the process of using it, therefore, no personal data whatsoever was used during this process. Additionally, all student participants were informed that their design work and participation in the study could be used for academic publication. Consent from them was obtained prior to the inclusion of their work, and participation was entirely voluntary. No personally identifiable information has been published.

**Informed Consent Statement:** Informed consent was obtained from all subjects involved in the study.

**Data Availability Statement:** The data presented in this study are available on request from the corresponding author. Data available on request due to privacy.

**Conflicts of Interest:** The author declares no conflict of interest.